\documentclass[fleqn,11pt]{olplainarticle}

\title{Round-robin differential phase-time-shifting protocol for quantum key distribution: theory and experiment}

\author[1,2]{Kai Wang}
\author[3,4]{Ilaria Vagniluca}
\author[1]{Jie Zhang}
\author[2]{Søren Forchhammer}
\author[3,5]{Alessandro Zavatta}
\author[6]{Jesper B. Christensen}
\author[2,*]{Davide Bacco}
\affil[1]{The State Key Lab of Information Photonics and Optical Communications, Beijing University of Posts and Telecommunications, Beijing, 100876, CH}
\affil[2]{Department of Photonics Engineering, Technical University of Denmark,  2800 Kgs. Lyngby, DK}
\affil[3]{Consiglio Nazionale delle Ricerche - Istituto Nazionale di Ottica (CNR-INO), 50125 Florence, IT}
\affil[4]{Department of Physics “Ettore Pancini", University of Naples “Federico II", 80126 Naples, IT}
\affil[5]{LENS and Department of Physics, University of Florence, 50019 Sesto Fiorentino, IT}
\affil[6]{Danish Fundamental Metrology, Kogle All\'e 5, 2970 H{\o}rsholm, DK}
\affil[*]{Corresponding author: dabac@fotonik.dtu.dk}


\begin{abstract}
Quantum key distribution (QKD) allows the establishment of common cryptographic keys among distant parties. Many of the QKD protocols that were introduced in the past, involve the challenge of monitoring the signal disturbance over the communication line, in order to evaluate the information leakage to a potential eavesdropper. Recently, a QKD protocol that circumvents the need for monitoring signal disturbance, has been proposed and demonstrated in initial experiments. Here, we propose a new version of this so-called round-robin differential phase-shifting (RRDPS) protocol, in which both time and phase degrees-of-freedom are utilized to enlarge the Hilbert space dimensionality, without increasing experimental complexity or relaxing security assumptions. We derive the security proofs of the round-robin differential phase-time-shifting (RRDPTS) protocol in the collective attack scenario, and benchmark the new protocol against RRDPS for different experimental parameters. Furthermore, a proof-of-concept experiment of the RRDPTS protocol, using weak coherent pulses and decoy state method, is demonstrated over 80 km of fiber link.  Our results show that the RRDPTS protocol can achieve higher secret key rate in comparison with the RRDPS, in the condition of high quantum bit error rate.  
\end{abstract}

\begin{document}

\flushbottom
\maketitle
\thispagestyle{empty}

\section*{Introduction}
The unconditional security of key agreement is the main challenge in encryption systems based on private keys. Based on the laws of quantum mechanics, quantum key distribution (QKD) allows information theoretic security \citep{pirandola2019,xu2020secure}, by enabling two remote parties to safely establish a secret key. Currently, three families of QKD protocols have been defined and experimentally validated: discrete variable (DV), continuous variable (CV) and distributed-phase reference (DPR) \citep{pirandola2019}. Each family exhibits several pro-and-cons and can be used  in different practical conditions and security frameworks. However, in most of these QKD protocols, the noise affecting the quantum communication is considered as eavesdropping activity. In fact, the information leakage to a malicious third party is usually estimated, and bounded, by the observed error rates over the quantum channel. As a consequence, the error rate has great influence on the secure key rate and on the transmission distance of the quantum protocols. In particular, QKD protocols have a limited tolerance for the errors, and when the error rate goes beyond a certain threshold level, a secure key cannot be generated. For example, the error rate threshold of a DV Bennet-Brassard 1984 protocol is 11$\%$ with one-way reconciliation technique \citep{bennett1984advances,li2016experimental}. Regarding the DPR family, a typical differential-phase-shifting (DPS) protocol can tolerate only up to 4\% \citep{hatakeyama2017differential}.

Recently, to overcome this problem, the round-robin differential-phase-shifting (RRDPS) protocol has been proposed \citep{sasaki2014practical} and experimentally tested \citep{takesue2015experimental,wang2015experimental,zhang2017practical,guan2015experimental}. In this protocol, the information leakage to a potential eavesdropper can be bounded only by the user's own settings, rather than the noise affecting the quantum communication. In other words, the RRDPS exhibits higher tolerance for the quantum bit error rate and allows the generation of a secure key, without monitoring the phase error rate for estimating the eavesdropping disturbances. In this way, the influence of statistical fluctuations, that typically arise in the monitoring process, can be circumvented. 

However, although a new and improved security bound for the RRDPS protocol has been presented by Z.-Q. Yin and co-authors in 2018 \citep{yin2018improved}, the final secret key rate achievable is still limited by the experimental conditions, that typically affect the stability of the apparatus. \textcolor{black}{In particular, an unstable visibility of interference can lead to a high error rate of quantum measurements, which limits the performance of the protocol.} Nevertheless, as demonstrated in previous works \citep{cozzolino2019high}, by exploiting the properties of high-dimensional quantum states it is possible to improve the photon-information efficiency and, at the same time, the robustness to noise. Therefore, in order to improve the secret key rate of RRDPS  \textcolor{black}{protocol, by enhancing} its tolerance with respect to the instability of the apparatus, we propose a round-robin differential phase-time-shifting (RRDPTS) protocol, in which the time of arrival and the relative phase of light pulses are used to encode two secret key bits in each $L$-pulse signal. This protocol is an extension of the DPTS protocol introduced by Bacco et al. in \citep{bacco2016two,da2019experimental}, belonging to the DPR family. The security analysis of the RRDPTS protocol under collective attacks is derived for different photon number statistics. In addition, in order to test our new proposal, a proof-of-concept experiment of decoy-state RRDPTS is performed over 80 km of fiber link. Our results show that, as compared with the RRDPS, the RRDPTS protocol can improve the secret key rate achievable in the case \textcolor{black}{of} high error rate. In general, the RRDPTS protocol exhibits high tolerance for extremely unstable visibility of the interferometer. For this reason, we believe that this protocol can pave the way towards a practical implementation of QKD, under unstable experimental environments. 

\begin{figure}[ht]
\centering
\includegraphics[width=0.87\linewidth]{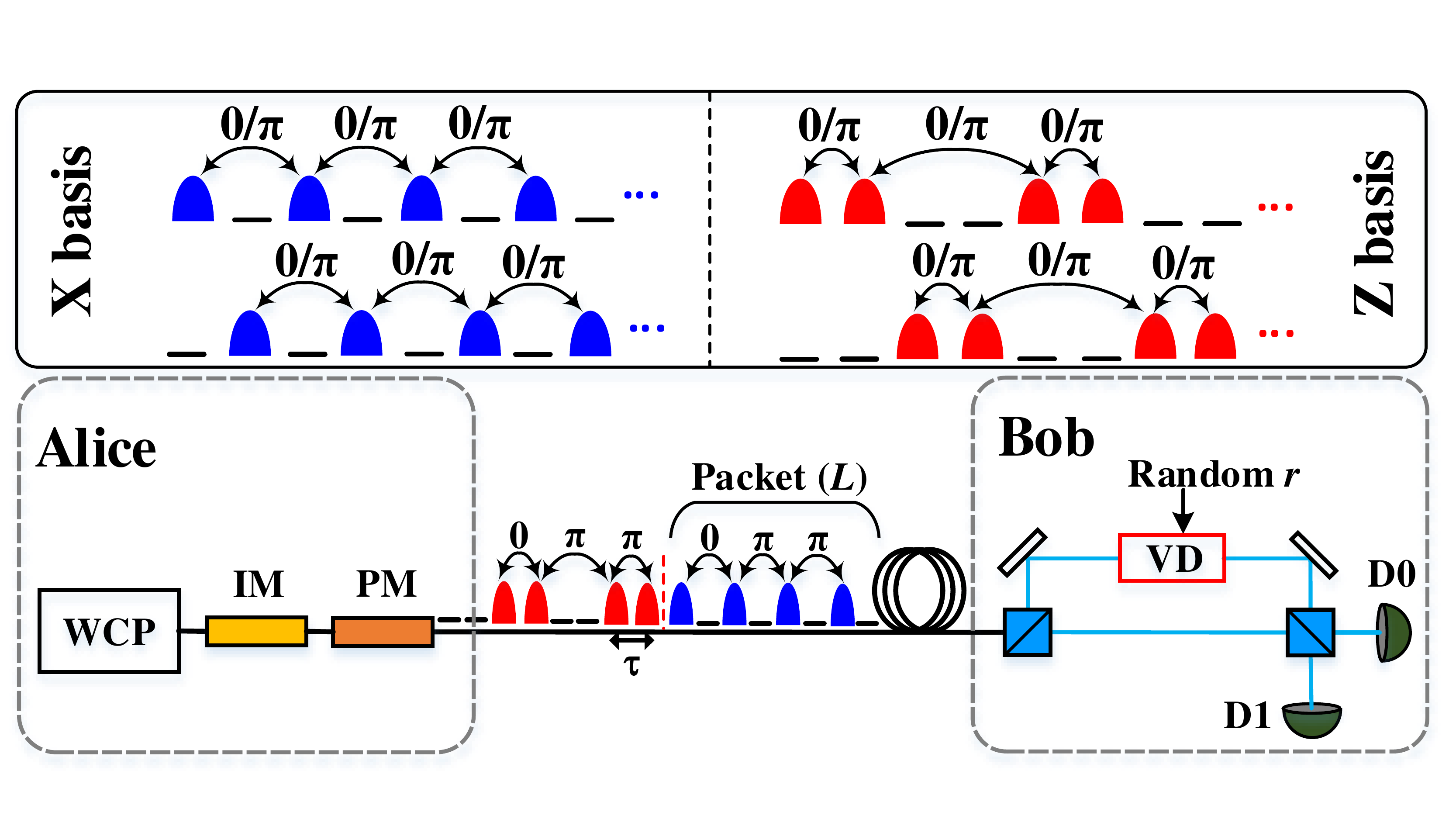}
\caption{\textbf{Schematic of the RRDPTS protocol}. We introduced two mutually unbiased bases (X basis and Z basis), in which different patterns of empty and non-empty time-bins are defined. Each of the $L$ pulses (i.e., non-empty bins) is prepared with a random relative phase $\left\{ 0, \pi\right\}$. In the bottom part of the figure, we depict the experimental devices necessary for running the RRDPTS protocol. Here, the exemplified pulse train illustrates the case of $L=4$. \textcolor{black}{WCP: weak coherent pulses;} PM: phase modulator; IM: intensity modulator; VD: variable delay; D0, D1: single-photon detectors.
}
\label{fig:RRDPTS0}
\end{figure}

\section*{Description of RRDPTS}
Since the RRDPTS protocol is included in the DPR family, the quantum states are encoded in subsequent pulses. In particular, by exploiting both the phase and time degrees of freedom of each $L$-pulse packet, the RRDPTS protocol can distribute two key bits per successfull detection event. The RRDPTS protocol is described in the following points:

(a) The sender (Alice) at first prepares packets containing $2L$ pulses, with repetition rate $1/\tau$ (where $\tau$ is defined as duration of a time bin). 
\textcolor{black}{Then,} the intensity of each $2L$-pulse packet is modulated to shape one of the two temporal profiles given by the X basis or the Z basis, depicted in Figure \ref{fig:RRDPTS0}. The two temporal bases, which are randomly chosen for each packet, differ by the time-bin positions occupied by the resulting $L$ pulses, and by the positions of the remaining $L$ bins that are left empty. In addition, a phase modulator is used to randomly modulate the relative phase ($0$ or $\pi$) of each pulse of the packet. In this way, a two-bit key is encoded in each packet containing the resulting $L$ pulses. With reference to Figure \ref{fig:RRDPTS0}, the states prepared at the transmitter are given by
 \begin{equation}\label{1}
  \centering
        \begin{cases}
            \left | X \right\rangle_0=\begin{matrix} \bigotimes_{k=1}^L \left | vac   \right \rangle_{2k-1} \left | \alpha e^{i\phi_{2k}}   \right \rangle_{2k} \end{matrix}  \\  
            \left | X \right\rangle_1=\begin{matrix} \bigotimes_{k=1}^L \left | \alpha e^{i\phi_{2k-1}}  \right \rangle_{2k-1} \left |  vac  \right \rangle_{2k} \end{matrix} 
 \end{cases}
\end{equation}    
 \begin{equation}\label{2}
  \centering
    \begin{cases}
             \left | Z \right\rangle_0=\begin{matrix} \bigotimes_{k=1}^{L/2} \left | \alpha e^{i\phi_{4k-3}} \right \rangle_{4k-3} \left | \alpha e^{i\phi_{4k-2}}  \right \rangle_{4k-2} \left | vac \right \rangle_{4k-1} \left | vac  \right \rangle_{4k} \end{matrix} \\ 
             \left | Z \right\rangle_1=\begin{matrix} \bigotimes_{k=1}^{L/2} \left | vac \right \rangle_{4k-3} \left | vac  \right \rangle_{4k-2} \left | \alpha e^{i\phi_{4k-1}} \right \rangle_{4k-1} \left | \alpha e^{i\phi_{4k}}  \right \rangle_{4k} \end{matrix}  
\end{cases}
\end{equation}  
where the subscript $\left\{0,1\right\}$ of $\left | X  \right \rangle_{0/1} (\left | Z  \right \rangle_{0/1}) $ refers to the bit encoded on the time-bin degree of freedom. Here, $\left | vac  \right \rangle $ is the vacuum state, while $\left |\alpha e^{i\phi_{2k}} \right\rangle_{2k}$ denotes a coherent state in the $2k$-th time-bin, with $\phi_{2k}$ phase and average photon number per pulse given by $\mu = \left | \alpha \right|^2$. The relative phase of each pulse is randomly selected between $\left\{ 0, \pi\right\}$\textcolor{black}{, with respect to the global phase of the packet (which is assumed to be phase-randomized)}. 

(b) Upon receiving each $L$-packet from the quantum channel, the receiver (Bob) sets the relative delay value $r\in\left\{1, \ 2, \ ...,\ L-1\right\}$, between the two paths of his adjustable Mach-Zehnder interferometer, as shown in Figure \ref{fig:RRDPTS0}. In particular, he randomly selects to measure the X basis, by setting a random delay $2r\tau\in\left\{2\tau, \ 4\tau, \ ..., \ (2L-2)\tau\right\}$, or to measure the Z basis, by setting a random delay $(2r-1)\tau\in\left\{ \tau, \ 3\tau, \ ..., \ (2L-3)\tau\right\}$. Then, Bob records the events when only one click occurs in his observed time-window. For each event, Bob extracts the time-encoded information by observing the time-bin at which the detection occurs, then he extracts the phase-encoded information by checking which one of the two detectors has clicked. 

(c) Alice and Bob repeat the above process and then publicly disclose the temporal basis that was selected for each event. Also, similarly to the RRDPS protocol, Bob announces the indices $\{a,b\}$ (with $a<b$ and $a,b\in\left\{1, \ 2, \ ...,\ L\right\}$) of the two pulses in the packet that have interfered with each other, thus enabling Alice to recover the relative phase information. Notably, Bob does not disclose the corresponding time-bin positions (1, 2, ..., $2L$), in order to keep secret the time-encoded information. Then, by collecting the events corresponding to the same choice of temporal basis, Alice and Bob can finally obtain the two-bit sifted key.

(d) Finally, Alice and Bob perform standard error correction procedure and privacy amplification \citep{buttler2003fast,bennett1995generalized}, in order to share a secure identical key.

Similar\textcolor{black}{ly} to the RRDPS protocol, the RRDPTS is robust against the intercept-resend attack, because of the information causality related to the random delay choice at the receiver \citep{sasaki2014practical}. 
However, the eavesdropper (Eve) can obtain the temporal information from the multi-photon packets, by means of photon-number splitting (PNS) attack \citep{brassard2000limitations,lutkenhaus2000security}.  To cope with this situation, Eve's information on multi-photon packets has to be bounded, in the case of weak coherent pulses (as we show in the following section). Furthermore, for long-distance QKD, decoy states with different intensities can be prepared in order to counteract the PNS attack (as we discuss in the Supplementary) \citep{hwang2003quantum,ma2005practical}.

\section*{Security Analysis of RRDPTS under Collective Attacks}
In order to derive the security proof for the RRDPTS protocol, we take inspiration from the work on RRDPS of Z.-Q. Yin and co-authors \citep{yin2018improved}. 
Here, we present the security proof of the RRDPTS protocol under collective attacks, using single-photon quantum states. In addition, we derive the security of the protocol using weak coherent pulses, under the general $N$-photon case (see Supplementary B-G). \\
In a collective attack scenario, Eve can probe the signal states with an ancilla and subsequently store the ancilla in a quantum memory. Then, after Bob has announced his measurement results to Alice, Eve performs a collective measurement of her ancilla states. We assume that Alice randomly prepares single-photon states from the X basis or the Z basis for each round of communication. In this section, we present the derivation of Eve's density matrix in the X basis (see Supplementary A for Eve's density matrix in the Z basis). The single-photon states from the X basis are given as follows
\begin{equation}\label{3}
\centering
    \begin{cases}
        \left | X \right\rangle_0=\frac{1}{\sqrt{L}}\begin{matrix} \sum_{i=1}^L  (-1)^{k_{2i}} \left | 2i  \right \rangle \end{matrix},  \\  
        \left | X \right\rangle_1=\frac{1}{\sqrt{L}}\begin{matrix} \sum_{i=1}^L (-1)^{k_{2i-1}} \left | 2i-1 \right \rangle \end{matrix}, 
    \end{cases}
\end{equation}   
where, $k_{2i}, \ k_{2i-1} \in \left\{ 0, 1\right\}$ carry the phase information, while the states $\left |2i \right\rangle$, $\left |2i-1 \right\rangle$ denote a single photon in the $2i$-th and $(2i-1)$-th time-bin positions, respectively ($i\in\left\{1,...,L\right\}$). \\
Eve's general collective attack on the X basis states can be  given by
\begin{equation}\label{4}
\centering
    \begin{cases}
         U_{Eve}\left | 2i \right\rangle \left | e_{00} \right\rangle = \begin{matrix} \sum_{j=1}^{L} c_{2i,2j} \left | 2j \right\rangle \left | e_{2i,2j} \right\rangle \end{matrix},\\
        U_{Eve}\left | 2i-1 \right\rangle \left | e_{00} \right\rangle = \begin{matrix} \sum_{j=1}^{L} c_{2i-1,2j-1} \left | 2j-1 \right\rangle \left | e_{2i-1,2j-1} \right\rangle \end{matrix},\\
    \end{cases}
\end{equation}
where, the subscript $\left\{2i,2j\right\}$ represents the shift from the $2i$-th position to the $2j$-th position, and $ \left | e_{2i,2j} \right\rangle $ is the corresponding quantum state of Eve's ancilla.\\ 
For each round of communication, Eve retains only her ancilla and sends Alice's photon to Bob. A successful event takes place when Bob projects the incoming state $\left|X\right\rangle_0$ or $\left|X\right\rangle_1$ onto $(\left|2a\right\rangle \pm \left|2b\right\rangle)/\sqrt{2}$ and $(\left|2a-1\right\rangle \pm \left|2b-1\right\rangle)/\sqrt{2}$, respectively. Then, the evolution of quantum states can be written as 
\begin{equation}\label{5}
\centering
    \begin{aligned}
        \left | X \right\rangle_0 \left | e_{00} \right\rangle  \to & \ U_{Eve}\left | X \right\rangle_0 \left | e_{00} \right\rangle\\
        &= \frac{1}{\sqrt{L}}\begin{matrix}\sum_{i=1}^{L}(-1)^{k_{2i}}(\tilde{c}_{2i,2a}\left|2a\right\rangle +\tilde{c}_{2i,2b}\left|2b\right\rangle)\end{matrix},
\end{aligned}
\end{equation}
\begin{equation}\label{6}
\centering
    \begin{aligned}
        \left | X \right\rangle_1 \left | e_{00} \right\rangle \to & \ U_{Eve}\left | X \right\rangle_1 \left | e_{00} \right\rangle\\
        & = \frac{1}{\sqrt{L}}\begin{matrix}\sum_{i=1}^{L}(-1)^{k_{2i-1}}(\tilde{c}_{2i-1,2a-1}\left|2a-1\right\rangle +\tilde{c}_{2i-1,2b-1}\left|2b-1\right\rangle)\end{matrix},
    \end{aligned}
\end{equation}
where, $\tilde{c}_{2i,2j}\triangleq c_{2i,2j}\left|e_{2i,2j}\right\rangle$ and $\tilde{c}_{2i-1,2j-1}\triangleq c_{2i-1,2j-1}\left|e_{2i-1,2j-1}\right\rangle$.\\
Consequently, the density matrix (non-normalized) of Eve's ancilla can be computed as the partial trace from the state above, giving
\begin{equation}\label{7}
\centering
    \begin{aligned}
        \rho_{x_0} = P \left\{ \sum_{i=1}^L (-1)^{k_{2i}} \tilde{c}_{2i,2a} \right\} + P \left\{ \sum_{i=1}^L (-1)^{k_{2i}} \tilde{c}_{2i,2b} \right\} ,
    \end{aligned}
\end{equation}
\begin{equation}\label{8}
\centering
    \begin{aligned}
        \rho_{x_1} = P \left\{ \sum_{i=1}^{L} (-1)^{k_{2i-1}} \tilde{c}_{2i-1,2a-1} \right\} + P \left\{ \sum_{i=1}^{L} (-1)^{k_{2i-1}} \tilde{c}_{2i-1,2b-1} \right\} ,
    \end{aligned}
\end{equation}
where, $P\left\{\left|x\right\rangle\right\}=\left|x\right\rangle\left\langle x\right|$ \textcolor{black}{and $\tilde{c}_{2i,2a}$, $\tilde{c}_{2i,2b}$, $\tilde{c}_{2i-1,2a-1}$, $\tilde{c}_{2i-1,2b-1}$ are the ket-vectors as defined above}. \\
After carrying out his projective measurement, Bob then announces his temporal basis choice and the indices $\left\{a,b\right\}$ (with $a<b$ and $a, \ b\in\left\{1, \ 2, \ ...,\ L\right\}$) of the interfering non-empty bins, over a public channel. Based on this announcement, Bob and Alice distill the phase and temporal information as raw key bits. In particular, the phase-encoded bit that Eve aims to guess, is given by $k_{2a}\oplus k_{2b}$ for $\left|X\right\rangle_0$ and by $k_{2a-1}\oplus k_{2b-1}$ for $\left|X\right\rangle_1$. So, Eve only cares about the relative phases between $\left\{2a,2b\right\}$ or $\left\{2a-1,2b-1\right\}$ positions, and ignores the phase in the other positions. 
Since for any $k_{2i}$ or $k_{2i-1}$ the phase $(-1)^{k_{2i}}$ or $(-1)^{k_{2i-1}}$ is completely random, the relative phase between $\left| e_{2a,2a}\right\rangle$ and $\left| e_{2i,2a}\right\rangle $ (or $\left| e_{2b,2b}\right\rangle$ and $\left| e_{2i,2b}\right\rangle$), with $i\ne a,b$, will be randomized. The same occurs between $\left| e_{2a-1,2a-1}\right\rangle,\left| e_{2b-1,2b-1}\right\rangle$ and $\left| e_{2i-1,2a-1}\right\rangle$,$\left| e_{2i-1,2b-1}\right\rangle$. Therefore, these components do not give Eve any information about the phase (although they may still leak some time-bin information to Eve). Thus, in order to access both the phase and temporal encoded information on quantum states, only the following components of the density matrix can be taken into account:
\begin{equation}\label{9}
\centering
    \begin{aligned}
        \rho_{x_0} \to & \ P\left\{ (-1)^{k_{2a}}\tilde{c}_{2a,2a} + (-1)^{k_{2b}}\tilde{c}_{2b,2a} \right\}
        + P\left\{ (-1)^{k_{2b}}\tilde{c}_{2b,2b} + (-1)^{k_{2a}}\tilde{c}_{2a,2b} \right\} \\
        & + \sum_{i \ne a, b} \Bigl( c_{2i,2a}^2 P\left\{\left | e_{2i,2a} \right\rangle\right\}
        + c_{2i,2b}^2 P\left\{\left | e_{2i,2b} \right\rangle\right\} \Bigr) \ ,
    \end{aligned}
\end{equation}
\begin{equation}\label{10}
\centering
    \begin{aligned}
        \rho_{x_1} \to & \ P\left\{ (-1)^{k_{2a-1}}\tilde{c}_{2a-1,2a-1} + (-1)^{k_{2b-1}}\tilde{c}_{2b-1,2a-1} \right\}
        + P\left\{ (-1)^{k_{2b-1}}\tilde{c}_{2b-1,2b-1} + (-1)^{k_{2a-1}}\tilde{c}_{2a-1,2b-1} \right\} \\
        & + \sum_{i \ne a, b} \Bigl( c_{2i-1,2a-1}^2 P\left\{\left | e_{2i-1,2a-1} \right\rangle\right\}
        + c_{2i-1,2b-1}^2 P\left\{\left | e_{2i-1,2b-1} \right\rangle\right\} \Bigr) \ .
    \end{aligned}
\end{equation}
As a result, when the relative phase between the $a$-th and $b$-th non-empty bins in the packet is 0 (i.e., $k_{2a}\oplus k_{2b}=0$ or $k_{2a-1}\oplus k_{2b-1}=0$), Eve's ancilla is written as 
\begin{equation}\label{11}
\centering
    \begin{aligned}
        \rho_{x_0,0} \to & \ P\left\{\tilde{c}_{2a,2a} + \tilde{c}_{2b,2a}\right\} 
        + P\left\{\tilde{c}_{2b,2b} + \tilde{c}_{2a,2b}\right\} 
        + \sum_{i \ne a, b} \Bigl( c_{2i,2a}^2 P\left\{\left | e_{2i,2a} \right\rangle\right\}
        + c_{2i,2b}^2 P\left\{\left | e_{2i,2b} \right\rangle\right\} \Bigr) \ ,
    \end{aligned}
\end{equation}
otherwise, when the relative phase is 1, Eve's ancilla is given by 
\begin{equation}\label{12}
    \centering
        \begin{aligned}
        \rho_{x_0,1} \to & \ P\left\{\tilde{c}_{2a,2a} - \tilde{c}_{2b,2a}\right\} 
        + P\left\{\tilde{c}_{2b,2b} - \tilde{c}_{2a,2b}\right\}  
        + \sum_{i \ne a, b} \Bigl( c_{2i,2a}^2 P\left\{\left | e_{2i,2a} \right\rangle\right\}
        + c_{2i,2b}^2 P\left\{\left | e_{2i,2b} \right\rangle\right\} \Bigr) \ ,
    \end{aligned}
\end{equation}
with $\rho_{x_1,0}$ and $\rho_{x_1,1}$ that can be written analogously. The corresponding derivation of the Z basis $\rho_{z_0,0}$, $\rho_{z_0,1}$ and $\rho_{z_1,0}$, $\rho_{z_1,1}$, is reported in Supplementary A.\\
Since Eve is only interested in distinguishing the two bits encoded on quantum states, and given that Alice chooses the X basis and Z basis randomly, then Eve's density matrix can be formulated in the following four cases:
\begin{equation}\label{13}
\centering
    \begin{cases}
        \rho_{(0,0)} = \frac{1}{2}( \rho_{x_0,0} +  \rho_{z_0,0} ) \\
        \rho_{(0,1)} = \frac{1}{2}( \rho_{x_0,1} +  \rho_{z_0,1} ) \\
        \rho_{(1,0)} = \frac{1}{2}( \rho_{x_1,0} +  \rho_{z_1,0} ) \\
        \rho_{(1,1)} = \frac{1}{2}( \rho_{x_1,1} +  \rho_{z_1,1} )  
    \end{cases}
\end{equation}
where, the first subscript of the density operator $\rho$ refers to the time-encoded bit, while the second subscript refers to the phase-encoded bit in the $a$-th and $b$-th pulses in the packet. Consequently, the information that Eve obtains on the raw key is given by (see Supplementary A for details)
\begin{equation}\label{15}
    \centering
            \begin{aligned}
                I_{AE}  
                & =  \frac{\begin{matrix} \sum_{a,b} Q^{(a,b)}I_{AE}^{(a,b)} \end{matrix}} {\begin{matrix} \sum_{a,b} Q^{(a,b)} \end{matrix}} 
                 \leq  \frac{f((L-1)x_1,x_2)+\frac{1}{8}(L-2)x_2+f(\frac{L/2-1}{2}y_1,\frac{1}{2}y_2) + \frac{1}{16}(L/2-2)y_2 } {\frac{1}{4}(L-1)(x_1 + x_2) + \frac{1}{4}(L/2-1)(y_1 + y_2)} \\
                & = \frac{f((L-1)x_1,x_2)+\frac{1}{8}(L-2)x_2+f(\frac{L/2-1}{2}y_1,\frac{1}{2}y_2) + \frac{1}{16}(L/2-2)y_2} {\frac{1}{2}(L-1)+\frac{1}{2}(L/2-1)}
            \end{aligned}
\end{equation}
where $f(x,y) = -\frac{x}{4}\log_4\frac{x}{4}-\frac{y}{4}\log_4\frac{y}{4}+\frac{x+y}{4}\log_4\frac{x+y}{2}$ and the non-negative real parameters $x_i$ and $y_i$ satisfy $ x_1 + x_2 = 2 $ and $ y_1 + y_2 = 2 $. Consequently, by using Eq. (\ref{15}), we can bound Eve's information for the RRDPTS protocol, without monitoring signal disturbances.\\
When considering a weak laser source instead of single photons, we assume to implement phase randomization \textcolor{black}{of the global phase of each packet}, in order to have a mixture of Fock states whose photon number follows a Poisson distribution. In this way, the security proof of RRDPTS protocol under general photon-number cases has been derived in Supplementary \citep{yin2018improved}. Our derivation shows that, for the RRDPTS protocol with $L$ pulses per packet and $N$ photon-number per packet, with $L/2\ge N+1$, Eve's information can be bounded by
\begin{equation}\label{16}
    \centering
       \begin{aligned}
            I_{AE} &\leq I_{AE}^U \\
            & \triangleq max_{x_1,...,x_{N+1},y_1,...,y_{N+1}}\left\{\frac{\sum_{n=1}^{N}f((L-n)x_n,nx_{n+1})+\frac{(L-N-1)x_{N+1}}{8}+\sum_{n=1}^{N}f(\frac{L/2-n}{2}y_n,\frac{n}{2}y_{n+1})+\frac{(L/2-N-1)y_{N+1}}{16}}{\frac{1}{2}((L-1)+(L/2-1))}\right\} \ .
       \end{aligned}
\end{equation}
with the non-negative real parameters $x_i$ and $y_i$ satisfying $\begin{matrix}\sum_{i=1}^{N+1} x_i = 2\end{matrix}$ and $\begin{matrix}\sum_{i=1}^{N+1} y_i = 2\end{matrix}$.\\
As demonstrated in the Supplementary, $I_{AE}<1$ always holds if the photon-number per packet satisfies $N\leq L/2-1$. Moreover, the upper bound for $I_{AE}$ is obtained without monitoring signal disturbance, which means that a more tight value for $I_{AE}^U$ can be established by observing the error rate introduced by Eve.  Furthermore, based on the quantum de Finetti theorem \citep{caves2002unknown,christandl2009postselection},  it is possible to demonstrate the security of RRDPTS protocol against coherent attacks. \\
Our derivation of $I_{AE}$ takes into account that, in RRDPTS protocol, Eve obtains also the temporal information when she gets access to the phase information from a given signal state. Moreover, some temporal information (but not the phase information) may leak out from the other components of the density matrix. The phase information leakage can be reduced by setting a larger $L$ \citep{sasaki2014practical}. However, in order to reduce the leakage of temporal information only, the photon-number per packet has to be decreased.  As a consequence, the threshold level of the photon-number per packet in RRDPTS protocol ($N\leq L/2-1$) is lower than in the RRDPS ($N\leq L-1$) \citep{yin2018improved}.

\section*{Results}
By bounding Eve's information under the general $N$-photon case, as reported in the previous section, we are now able to derive the secret key rate of the RRDPTS protocol, with $L$ as the number of pulses \textcolor{black}{(i.e., non-empty bins)} per packet:
\begin{equation}\label{38}
\begin{aligned}
           L \cdot R &= 2(Q(1-H(A|B))-e_{src}-(Q-e_{src})I_{AE}^U) \\
                     & = \textcolor{black}{2Q(1-H(A|B)-\frac{e_{src}}{Q}-(1-\frac{e_{src}}{Q})I_{AE}^U)} 
\end{aligned}
\end{equation}
where, $ R $ is the secret key rate per pulse, $Q$ is the probability to have only one click per packet, $ H(A|B) $ is the conditional entropy and $ e_{src}=1-\frac{1}{2}\sum_{n=0}^{v_{th}}(e^{-L\mu}(L\mu)^n/n!+e^{-L\mu/2}(L\mu/2)^n/n!)$ is the probability that the photon-number per packet is greater than $ v_{th} = N $. Here, $\mu$ is the mean photon number per pulse. The pre-factor 2 in Eq. (\ref{38}) is added since two bits of raw key are generated for each successful detection event.
In order to calculate $ Q $, we assume a lossy channel with $t$ transmission, while $ \eta_d $ is the detection efficiency of single-photon detectors, with $ p_d $ dark-count probability per time bin.\\
The probability $ Q $ is associated with the temporal basis chosen by Bob. To select the X basis, Bob sets a temporal delay $ 2r\tau $, with the $ r $ value that is chosen at random between $ \left\{1,2,...,L-1\right\} $. Depending on the $ r $ value, both of Bob's detectors will open $ 2(L-r) $ time windows of $ \tau $ duration, in order to detect the interference of the incoming signal states. Considering that a click may arise from a signal state or a dark count, the probability that Bob obtains a single click per packet among the observed $2(L-r)$ windows, can be computed as \citep{yin2018improved}:
\begin{equation}\label{18}
    \centering
         Q_{x,r} = (1-p_d)^{4(L-r)-1}e^{-(L-r)\eta\mu}(L-r)(\frac{1}{2}\eta\mu+4p_d)
\end{equation}
where, $\eta=t \cdot \eta_d$. Here, the factor $1/2$ is added because half of the time bins are empty and invalid for distilling key bits. Accordingly, the mean yield of a single click per packet in the X basis is given by $Q_x = \begin{matrix} \sum_{r=1}^{L-1} Q_{x,r}/(L-1) \end{matrix}$. With similar reasoning, we can derive the contributions to the symbol error rate in the X basis:
\begin{equation}\label{20}
\centering
    \centering
         E_{x,r}^{(I)} Q_{x,r} = (1-p_d)^{4(L-r)-1}e^{-(L-r)\eta\mu}(L-r)(\eta\mu\frac{1}{2} e_{mis}+p_d)
\end{equation}
\begin{equation}\label{21}
\centering
    \centering
         E_{x,r}^{(II)} Q_{x,r} = E_{x,r}^{(III)} Q_{x,r} =(1-p_d)^{4(L-r)-1}e^{-(L-r)\eta\mu}(L-r)p_d
\end{equation}
where $E_{x,r}^{(I)} Q_{x,r}$ is the contribution of phase errors, that depends on the misalignment $e_{mis}$ of the interferometer, related to the intrinsic visibility of interference $V$ ($e_{mis}=\frac{1-V}{2}$). The other two contributions refer to the occurrence of a dark count generating time errors ($E_{x,r}^{(II)} Q_{x,r}$) or phase and time errors ($E_{x,r}^{(III)} Q_{x,r}$), and do not depend on $e_{mis}$, since the interferometer misalignment does not alter the measurement of arrival time. Consequently, as the the time-encoded information remains correct, the RRDPTS protocol suffers less from interferometer imperfections than the RRDPS protocol, which solely relies on the relative phase measurements.\\
The mean symbol error rate is given by $E_{x}^{(i)} Q_{x}=\sum_{r=1}^{L-1}E_{x,r}^{(i)} Q_{x,r}/(L-1)$, with $i\in\left\{I,II,III\right\}$.\\
Otherwise, to measure the Z basis, Bob sets a random delay $(2r-1)\tau$ (with $r\in \left\{2,...,L-1\right\}$). In order to evaluate $Q_{z,r}$, we have to distinguish between the even and odd values of $r$. For even values of $r$, 
both of Bob's detectors will open $ 2(L/2 -r/2)$ time windows to detect the phase and temporal information. Thus, the click-per-packet probability and the symbol error rates in the Z basis, for even values $r$, are given by
\begin{equation}\label{22}
    \centering
        Q_{z,r,e} = (1-p_d)^{2L-2r-1}e^{-(L/2-r/2)\eta\mu}(L/2-r/2)(\eta\mu\frac{1}{2}+4p_d) \ ,
\end{equation}
  \begin{equation}\label{25}
    \centering
        E_{z,r,e}^{(I)}Q_{z,r,e} = (1-p_d)^{2L-2r-1}e^{-(L/2-r/2)\eta\mu}(L/2-r/2)(\eta\mu\frac{1}{2} e_{mis}+p_d) \ , 
 \end{equation}
  \begin{equation}\label{26}
    \centering
        E_{z,r,e}^{(II)}Q_{z,r,e} = E_{z,r,e}^{(III)}Q_{z,r,e} = (1-p_d)^{2L-2r-1}e^{-(L/2-r/2)\eta\mu}(L/2-r/2)p_d \ .
 \end{equation}
On the other hand, for odd values of $ r $, both of Bob's detectors will open $ 2(L/2 -(r-1)/2) $ time windows, and we will have
 \begin{equation}\label{27}
        Q_{z,r,o} = (1-p_d)^{2L-2r+1}e^{-(L/2-(r-1)/2)\eta\mu}(L/2-(r-1)/2)(\eta\mu\frac{1}{2}+4p_d) \ ,
 \end{equation}
  \begin{equation}\label{30}
        E_{z,r,o}^{(I)}Q_{z,r,o} = (1-p_d)^{2L-2r+1}e^{-(L/2-(r-1)/2)\eta\mu}(L/2-(r-1)/2)(\eta\mu\frac{1}{2} e_{mis}+p_d) \ ,
 \end{equation}
  \begin{equation}\label{31}
        E_{z,r,o}^{(II)}Q_{z,r,o} = E_{z,r,o}^{(III)}Q_{z,r,o} = (1-p_d)^{2L-2r+1}e^{-(L/2-(r-1)/2)\eta\mu}(L/2-(r-1)/2)p_d \ .
 \end{equation}
Accordingly, the mean click rate per packet and symbol error rates for the Z basis are $Q_{z,e} = \begin{matrix} \sum_{r=2,e} Q_{z,r,e}/(L/2-1) \end{matrix}$, $E_{z,e}^{(i)}Q_{z,e}=\sum_{r=2,e}E_{z,r,e}^{(i)}Q_{z,r,e}/(L/2-1)$ and $Q_{z,o} = \begin{matrix} \sum_{r=3,o} Q_{z,r,o}/(L/2-1) \end{matrix} $, $E_{z,o}^{(i)}Q_{z,o}=\sum_{r=3,o}E_{z,r,o}^{(i)}Q_{z,r,o}/(L/2-1)$ for even and odd values of $r$, respectively ($i\in\left\{I,II,III\right\}$). 
So, by combining the previous equations, the overall probability to have one click per packet is given by
 \begin{equation}\label{33}
    \centering
    \begin{aligned}
        & Q = \frac{1}{2} Q_x + \frac{1}{2} Q_z = \frac{1}{2} Q_x + \frac{1}{4} Q_{z,e} + \frac{1}{4} Q_{z,o} \ .
    \end{aligned}
\end{equation}
Furthermore, the conditional entropy $H(A|B)$ is expressed as
\begin{equation}\label{34}
    \centering
        H(A|B) = -(1-e)\log_4(1-e)-\sum_{i} e^{(i)}\log_4(e^{(i)}) \ ,
\end{equation}
where the error probabilities are given by $e^{(i)} = E_{x}^{(i)}/2 + E_{z,e}^{(i)}/4 + E_{z,o}^{(i)}/4 $ and $e=\begin{matrix} \sum_{i} e^{(i)}\end{matrix}$, with $i\in\left\{I,II,III\right\}$.\\
It has to be noted that, here, we discard the $r=1$ case for the Z basis. The reason of this is that, when $r=1$, the leaked information from multi-photon packet from Z basis can not be bounded in a similar way as usually done for round-robin protocols. This obviously holds for a weak laser source, but not in the single photon case (as shown in Supplementary A). Besides, when implementing the decoy-state method on a weak laser source, it is possible to bound the information leakage also in the $r=1$ case for Z basis (Supplementary H).

\begin{figure}[htbp]
\centering                                                           
\subfigure[$e_{mis}=0.015$]{                    
\begin{minipage}{9cm}
\centering                                                          
\includegraphics[scale=0.275]{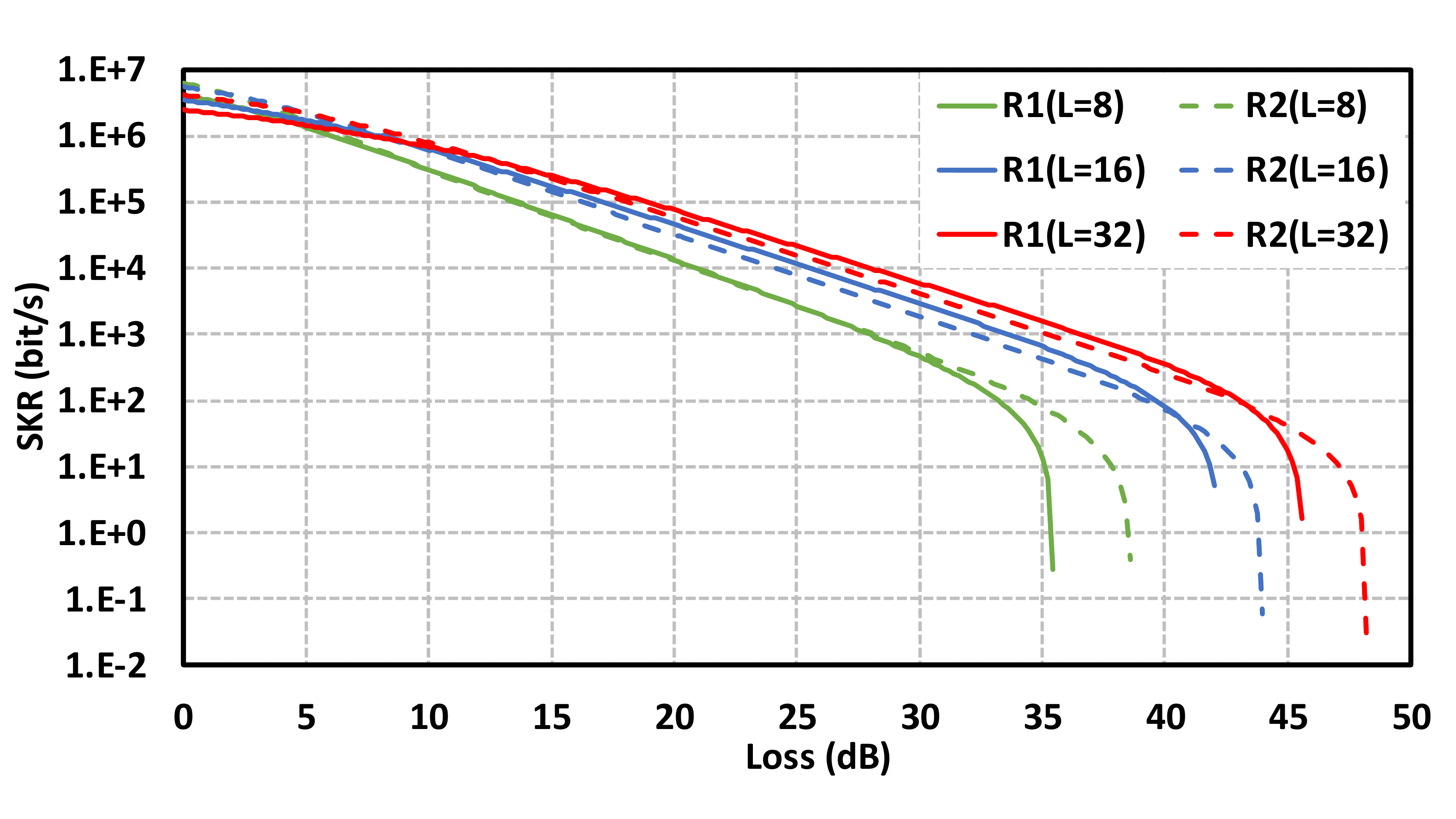}     
\end{minipage}
}
\subfigure[$e_{mis}=0.15$]{                  
\begin{minipage}{9cm}
\centering                                                         
\includegraphics[scale=0.275]{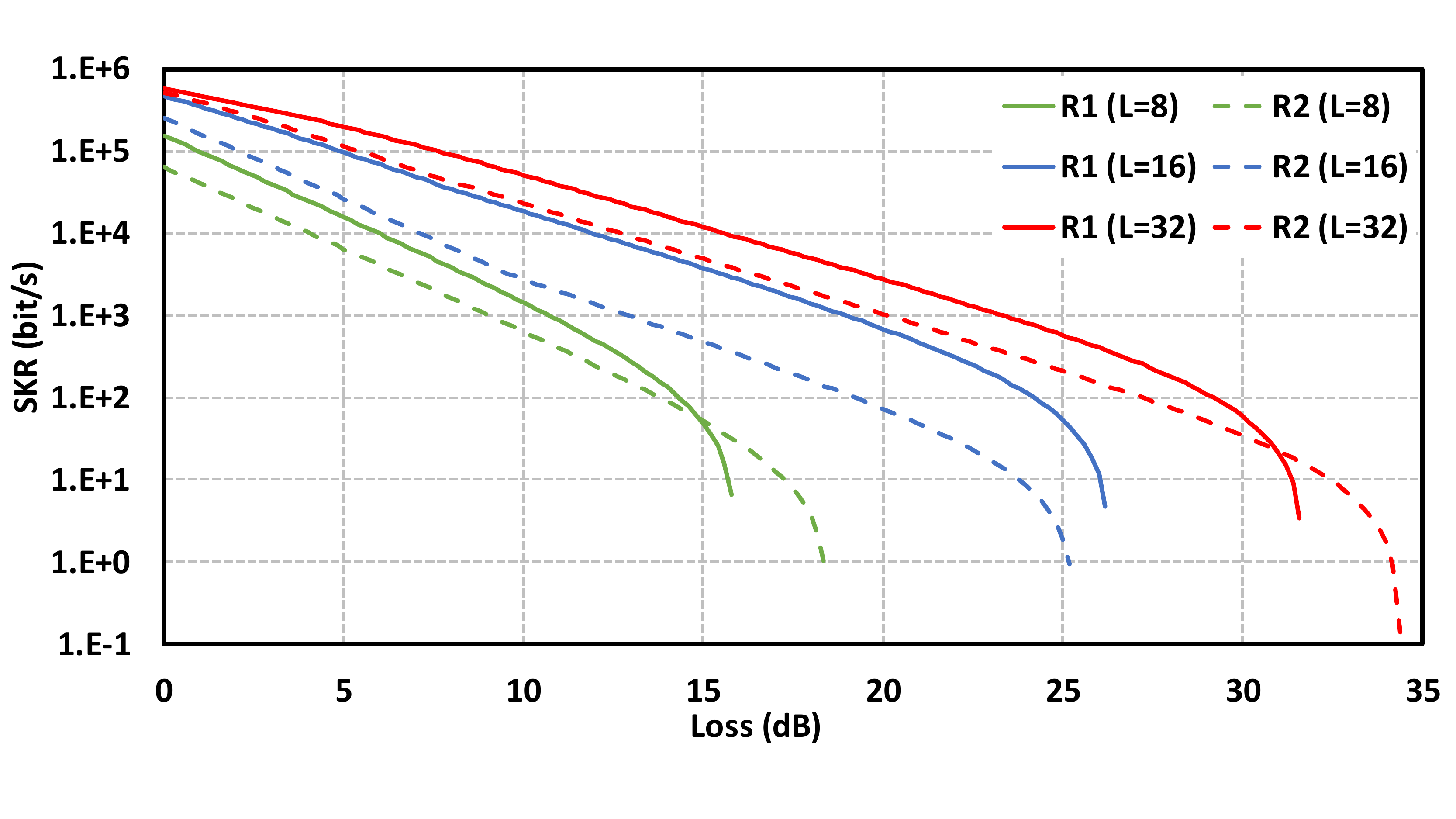}                
\end{minipage}
}
\caption{\textbf{\textcolor{black}{Secret key rate as a function of the channel loss, with two different visibility values and for three packet sizes.}} $R_1$ and $R_2$ represent the RRDPTS and RRDPS protocols, respectively, with weak coherent pulses. Both $R_1$ and $R_2$ are simulated in the scenarios without monitoring signal disturbance.
}              
\label{fig:2}                                                        
\end{figure}

\section*{Discussion}
Here, we present and discuss the secret key rate of the RRDPTS protocol, as a function of the achievable distance, and interference visibility, for different packet sizes ($L$). Furthermore, we compare the results of RRDPTS and RRDPS protocols, through numerical simulations. Both the protocols are simulated in the scenario without monitoring signal disturbance, with a weak coherent source at the transmitter\textcolor{black}{, exhibiting $1/\tau$ = 1 GHz repetition rate,} and superconductive detectors at the receiver, exhibiting $\eta_d = 85\%$ detection efficiency and $p_d = 1.6\times 10^{-8}$ dark count probability per time bin.\\
In Figure \ref{fig:2} we report the secret key rate, as a function of channel loss, of the RRDPTS and the RRDPS protocols ($R_1$ and $R_2$ respectively), for three different packet lengths ($L=8, \ 16, \ 32$) and two visibility values of Bob's interferometer ($e_{mis}=0.015$ and $e_{mis}=0.15$). Here, $R_1$ and $R_2$ are given by Eq. (\ref{38}) and reference \citep{yin2018improved}, respectively. The first consideration is related to the packet length. In the case of large packet size ($L=16$ and $L=32$, i.e., blue and red lines) and high visibility of interference ($e_{mis}=0.015$), the secret key rate of RRDPTS is slightly higher than that of RRDPS in low and middle loss regime, as shown in Figure \ref{fig:2}(a). Due to the higher dimensionality of encoding, the RRDPTS protocol can tolerate more information leakage $I_{AE}$ from multi-photon packets, as compared to RRDPS protocol. However, when the packet size $L$ decreases, this advantage will be offset, being $N\leq L/2-1$ for RRDPTS. Thus, in the case of a short packet length ($L=8$, green line), the RRDPTS protocol has similar performance with the RRDPS. 
Nonetheless, if we consider a lower visibility in the interferometer (see Figure \ref{fig:2}(b)), the RRDPTS protocol outperforms the RRDPS protocol. This is because the RRDPTS exploits also the time-bin encoding for carrying the secret bits, which is not influenced by the interference misalignment, as already shown in the symbol-error rate derivation of the previous section. Conversely, the RRDPS can rely only on the phase-encoded information, which is more likely affected by errors, as the visibility decreases. On the other hand, the fact that the RRDPTS exploits two-bit key per packet, makes it more sensitive to the dark counts affecting the detection. This is the reason why, for high channel loss (i.e., when the dark counts become relevant) the secret key rate of RRDPTS drops off more quickly than that of RRDPS. This behavior, shown in both Figure \ref{fig:2}(a) and (b), is generally exhibited by QKD protocols with high-dimensional encoding \citep{cozzolino2019high,bacco2016two,vagniluca2020efficient}. As a consequence, for a fixed visibility of interference, the RRDPTS protocol is more suitable for deployment covering shorter distances.

\begin{figure}[htbp]
\centering                                                           
\subfigure[10 dB loss]{                    
\begin{minipage}{9cm}
\centering                                                          
\includegraphics[scale=0.275]{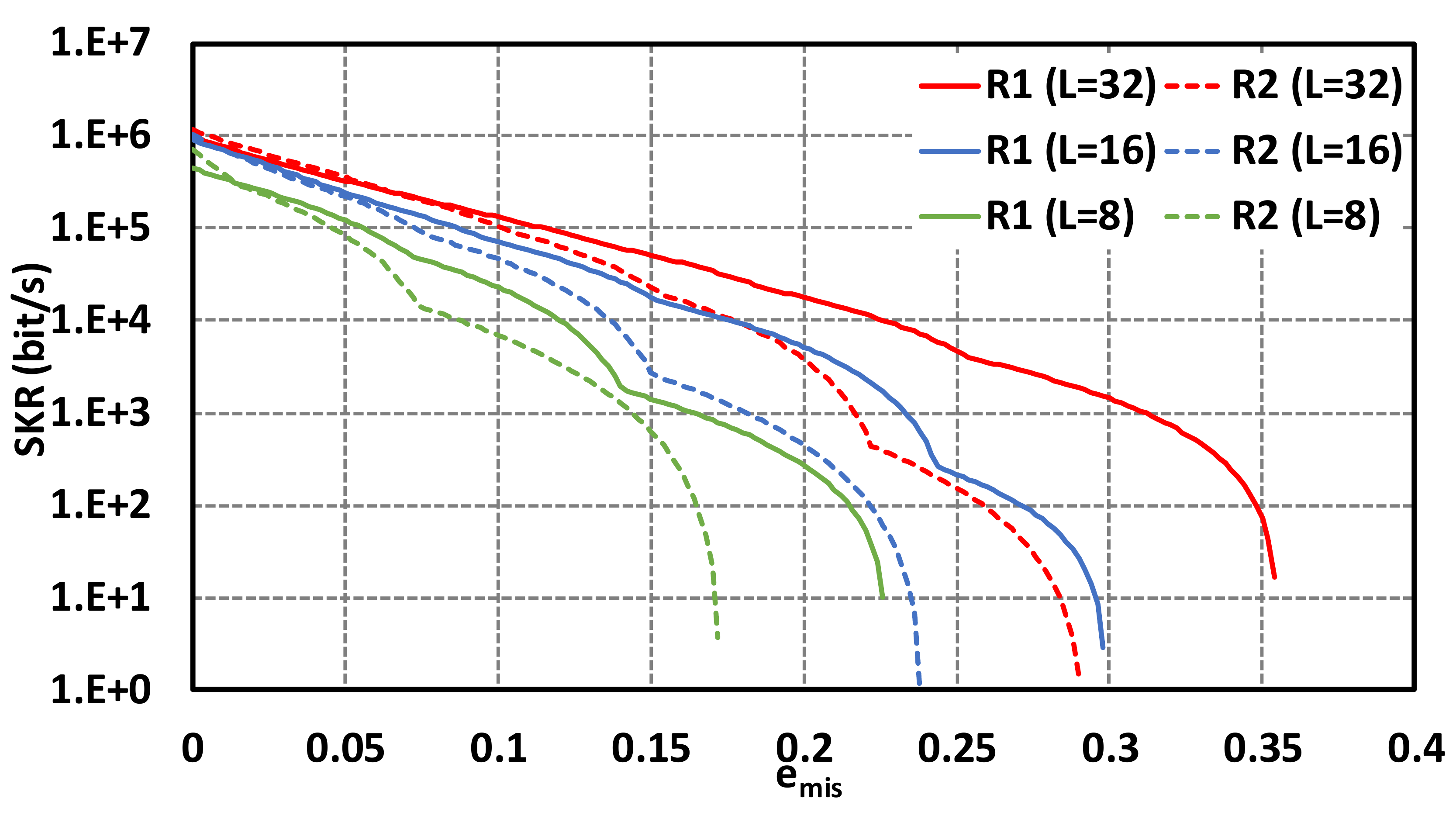}     
\end{minipage}
}
\subfigure[20 dB loss]{                  
\begin{minipage}{9cm}
\centering                                                         
\includegraphics[scale=0.275]{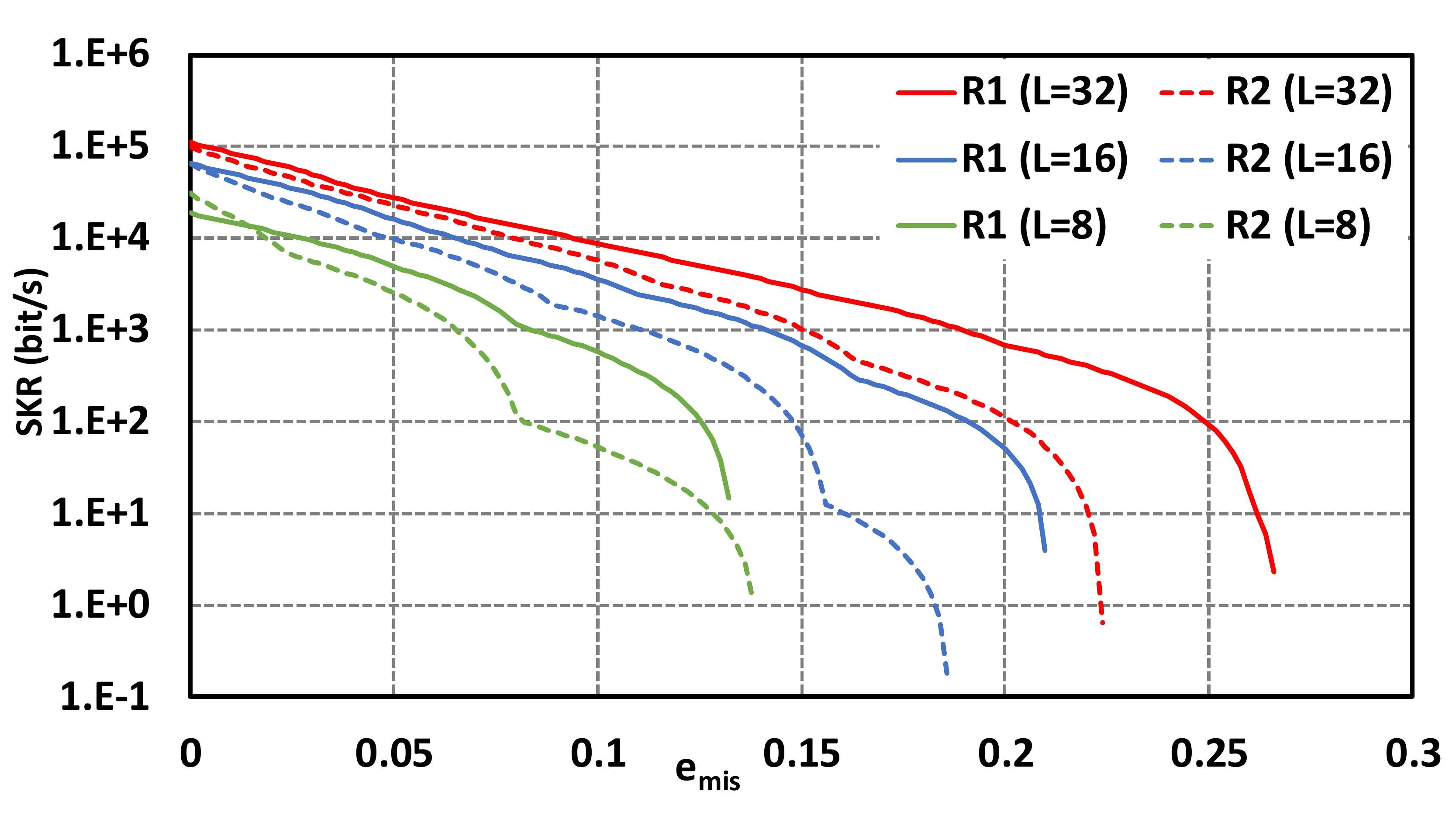}                
\end{minipage}
}
\caption{\textbf{\textcolor{black}{Secret key rate as a function of the interference misalignment, with two different channel loss and for three packet sizes.}} $R_1$ and $R_2$ represent the RRDPTS and RRDPS protocols, respectively, with weak coherent pulses. Both $R_1$ and $R_2$ are simulated in the scenarios without monitoring signal disturbance.}
\label{fig:3}                                            
\end{figure}

Figure \ref{fig:3} shows the secret key rates $R_1$, $R_2$ as a function of the misalignment error ($e_{mis}$) for two different channel loss values. Notably, the plotted lines of the secret key rate in Figure \ref{fig:3} show that the secret key rate is maximized with $\nu_{th}$ for each loss value, resulting in the apparent piecewise behavior. 
Generally, as the interference visibility decreases, the secret key rate achievable by RRDPTS is higher than that of RRDPS. Besides, the difference between RRDPTS and RRDPS, in terms of secret key rate, decreases for longer transmission distance, as shown in Figure \ref{fig:3}(b). Most importantly, thanks to the higher tolerance for interference errors, the RRDPTS returns a positive secret key rate, even in the conditions when the interferometer misalignment prevents a secure key to be established by using RRDPS protocol. \\
Furthermore, it is worth noticing that the RRDPTS protocol with decoy states can improve the secret key rate achievable, especially at long transmission distances, as reported in the Supplementary Material (see Figure S 3(a) in Supplementary). This is because in the long distance scenario and in the case of low visibility, the multi-photon packets leak more information to Eve. As a result, the RRDPTS protocol with decoy states performs better than decoy-free RRDPTS.

\section*{Proof-of-principle Experiment}
\begin{figure}[ht]
\centering
\includegraphics[width=0.87\linewidth]{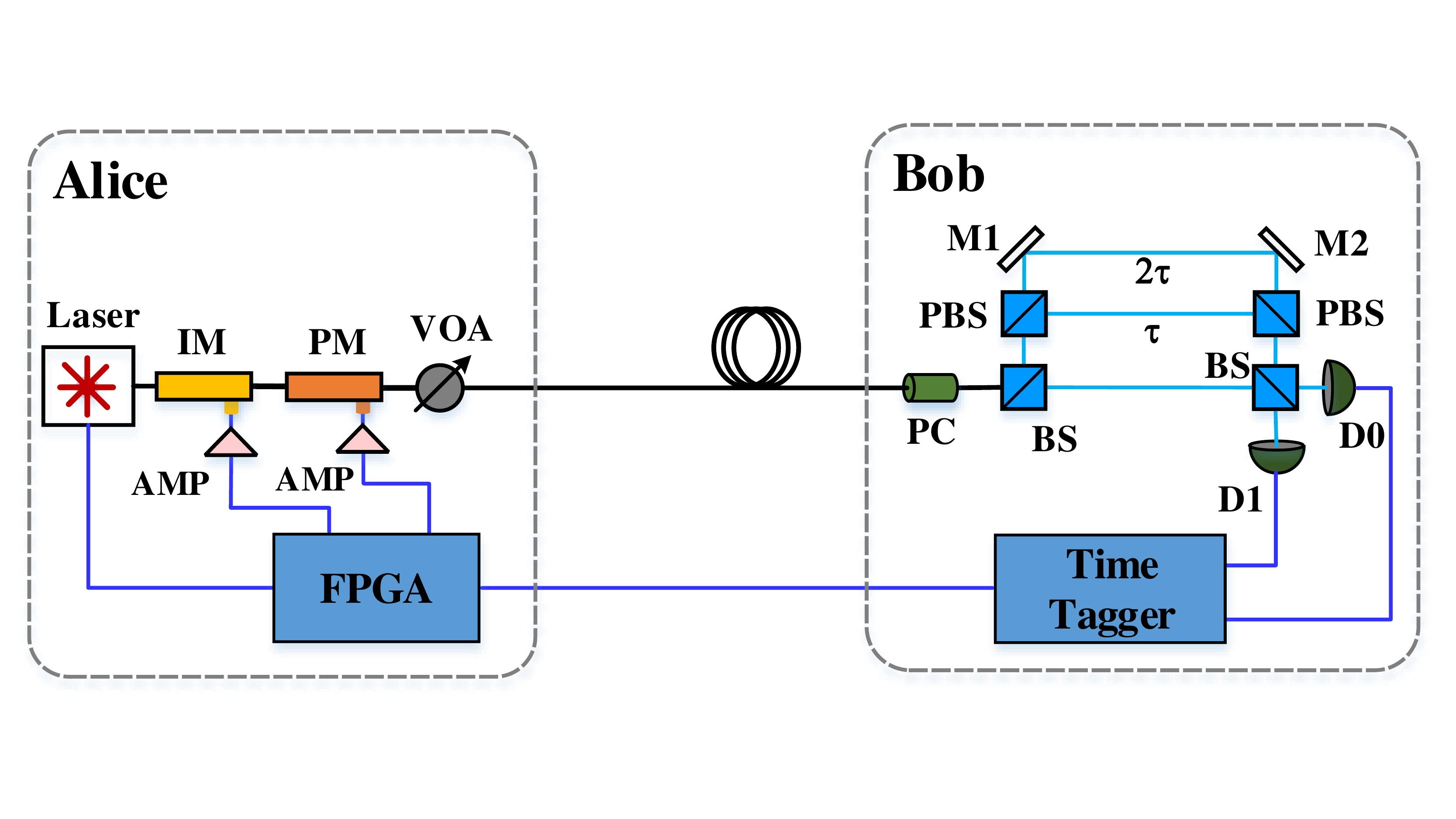}
\caption{\textbf{Setup of the proof-of-concept experiment}. Experimental setup of the RRDPTS protocol with $L=4$ and one-decoy method. Only two different delays ($\tau$ and $2\tau$) were tested in this proof-of-concept experiment, by means of two overlapping delay-line interferometers that were assembled at the receiver side (Bob). PM: phase modulator; IM: intensity modulator; AMP: RF amplifier; VOA: variable optical attenuator; FPGA: field programmable gate array; PC: polarization controller; BS: beam splitter; PBS: polarizing beam splitter; M1, M2: mirrors; VD: variable delay; D0, D1: InGaAs single-photon detectors.
}
\label{fig:RRDPTS7}
\end{figure}


\begin{figure}[htbp]
\centering                 
\subfigure[]{ 
\begin{minipage}{8.9cm}
\centering                                                         
\includegraphics[scale=0.6]{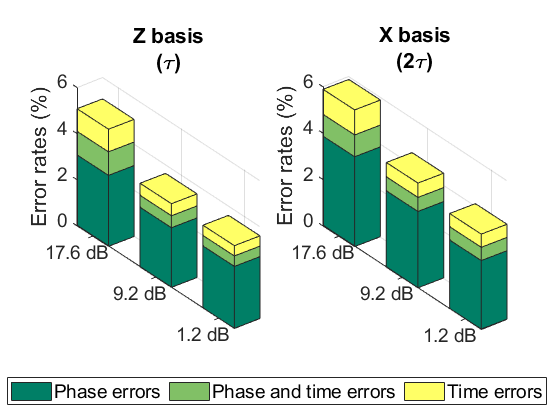}                
\end{minipage}
}
\subfigure[]{ 
\begin{minipage}{8.9cm}
\centering                                                          
\includegraphics[scale=0.6]{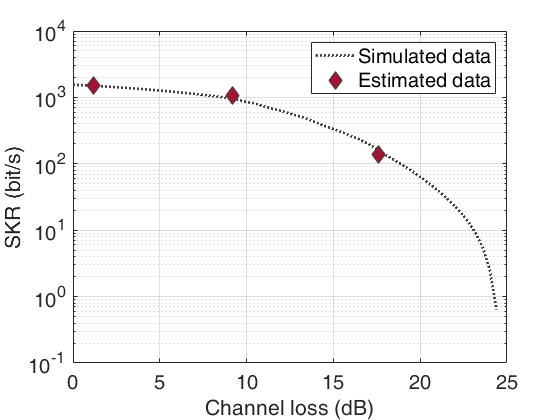}     
\end{minipage}
}
\caption{\textbf{\textcolor{black}{Results of the proof-of-concept experiment.}} Figure (a) shows the three different contributions to the symbol error rate that was measured in our experiment. Notably, the interference misalignment (phase errors) is generally the main contribution to the overall erroneous detections. \textcolor{black}{Figure (b)} shows the estimated secret key rate achievable by our experimental setup, as a function of channel loss. The dashed line represents the simulated behavior of one-decoy RRDPTS protocol with $L=4$, while red diamonds correspond to the expected secret key rate that we estimated from our experimental data.}
\label{fig:RRDPTS}                                            
\end{figure}

In order to test the properties of the new proposed protocol, a proof-of-concept experiment of RRDPTS is presented. Specifically, we tested the one-decoy RRDPTS scheme with $L=4$ and two different delays ($\tau$ and $2\tau$).\\
The experimental setup is depicted in Figure \ref{fig:RRDPTS7}. At the transmitter side (Alice), the two bases X and Z are prepared by applying sequential intensity and phase modulation to a continuous-wave laser source, emitting at 1550 nm. The time-bin duration is $\tau = 840$ ps, and two bits of information are encoded in each packet of eight time bins, consisting of four empty bins and four pulses with random $0$, $\pi$ relative phases. The mean photon number per pulse is adjusted with another intensity modulator and a variable optical attenuator. In particular, 50\% of the packets are prepared with $\mu$ intensity level (signal states) and the other 50\% with $\nu$ intensity level (decoy states). Then, the quantum states are sent into the fiber channel and reach the receiver apparatus (Bob). As shown in Figure \ref{fig:RRDPTS7}, when Bob chooses to measure the Z basis, the packet is sent to a $\tau$ delay-line, otherwise, to measure the X basis, the packet is sent to a $2\tau$ delay-line. The two different optical paths, which originate (and recombine) from a polarizing beam splitter\textcolor{black}{s} (PBS), are selected with a polarization controller that is placed in Bob's setup. This configuration yields to two independent interferometers, with $\tau$ and $2\tau$ delay respectively. The other possible delays ($3\tau$, $5\tau$ for Z basis and $4\tau$, $6\tau$ for X basis) are not tested in this proof-of-principle experiment, and are simulated only. The state preparation rate at the transmitter is approximately $72.7 \times 10^6$ packets per second, and the probability to choose Z or X basis is set to 50\% at both Alice's and Bob's sides. The two interferometer outputs are monitored with InGaAs single-photon detectors. The modulators at the transmitter are driven with a field programmable gate array (FPGA), which also provides a synchronization signal to the time tagging unit that collects the measurements at the receiver. \\
In table \ref{tab:exptab} we report the experimental parameters and results, collected by testing our setup with four different channel lengths of standard single-mode fiber. Here, $D_Z^{\tau}$ and $D_X^{2\tau}$, $E_Z^{\tau}$ and $E_X^{2\tau}$ are the detection rates (sifted) and the symbol error rates, respectively, measured for the two bases and including both $\mu$ and $\nu$ detections. As one can notice, for a fixed channel loss, $D_X^{2\tau}$ and $E_X^{2\tau}$ are always slightly higher as compared to $D_Z^{\tau}$ and $E_Z^{\tau}$, because of the different amount of $\tau$-windows that are observed for detecting the interference within each packet, for the two bases (six windows observed for $2\tau$ delay, four windows for $\tau$ delay). One one hand, a larger time-window leads to a higher sifted-key rate, on the other hand it also increases the probability of measuring a dark count, thus raising the error rate. Besides, we noticed that in both bases, the relative phase measurement gives always a major contribution to the symbol error rate, than the temporal bit measurement. Specifically, we experimentally evaluated the three different contributions to the overall symbol error rates $E_Z^{\tau}$ and $E_X^{2\tau}$, that are reported in Figure \ref{fig:RRDPTS}(a). This plot shows that, in all cases, phase errors (dark green) caused by interference misalignment, are the main contribution to the overall error rate, leading to around 70\% of all erroneous detections. By contrast, the other two contributions (yellow and light green) are roughly equal to each other (according\textcolor{black}{ly} to theory\textcolor{black}{, see Eqs. (\ref{21}),(\ref{26}),(\ref{31})}) and generate together only 30\% of all erroneous detections. In particular, only around 15\% of the overall symbol errors is caused by time errors, i.e., Bob mistaking $\left|X\right\rangle_0$ for $\left|X\right\rangle_1$ (or $\left|Z\right\rangle_0$ for $\left|Z\right\rangle_1$). This is due to the fact that, from an experimental point of view, retrieving the time-encoded bit is generally more straightforward that retrieving the phase-encoded bit, since the latter is influenced by the quality of interference (i.e., intrinsic visibility) while the former is not. By this way, as already pointed out in the previous sections, the additional temporal encoding gives the RRDPTS protocol a practical advantage over the RRDPS.\\
Figure \ref{fig:RRDPTS}(b) shows the simulated secret key rate (dashed line) achievable with one-decoy RRDPTS protocol \textcolor{black}{with $L=4$}, as a function of channel loss. Our simulation takes into account the practical limitations of our experimental setup, such as the finite dead time of single-photon detectors. The four dots correspond to the expected secret key rate values (reported also in Table \ref{tab:exptab}) that can be estimated from our experimental results. To compute the expected secret key rate, the other possible delays for Z basis and X basis were simulated from the collected data for $\tau$ and $2\tau$ delay, respectively. Specifically, detection and error rates were estimated from $D_Z^{\tau}$ ($D_X^{2\tau}$) and $E_Z^{\tau}$ ($E_X^{2\tau}$), by taking into account the different amount of time-windows that have to be observed for each different delay. Finally, the detailed derivation of the secret key rate of RRDPTS with decoy-state method is reported in Supplementary Material H-I.

\begin{table*}[ht]
\caption{{\bf Experimental parameters and results of the proof-of-concept experiment.} For each fiber channel that was tested, we report the mean photon number per pulse of signal ($\mu$) and decoy ($\nu$) states, the sifted detection rates ($D_Z^{\tau}$, $D_X^{2\tau}$) and symbol error rates ($E_Z^{\tau}$, $E_X^{2\tau}$) measured for Z basis (with $\tau$ delay) and X basis (with $2\tau$ delay) and including both $\mu, \ \nu$. Finally, the secret key rate (SKR) achievable by our setup, is estimated from the collected data.}
\begin{center}
\begin{tabular}{cc|ccccccc}
\hline\hline
&&&&&& \\[-0.6em]
\multicolumn{2}{c|} {{\bf fiber   channel}}  & \multicolumn{5}{c}{{\bf one-decoy RRDPTS ($L=4$)}} \\ [5pt] 
\hline
&&&&&& \\[-0.6em]
length  & loss    & $\mu$    & $\nu$   & $D_Z^{\tau}$ & $D_X^{2\tau}$ & $E_Z^{\tau}$ & $E_X^{2\tau}$ & SKR (estimated)   \\ [5pt] \hline
&&&&&& \\[-0.4em]
6 km   &  1.2 dB   & 0.020    & 0.010   &  7.7 kc/s  & 11.2 kc/s   & 3.6\%      &   4.1\%   & 1.5 kbit/s        \\ [7pt] 
43 km  & 9.2 dB    & 0.034    & 0.016   &  5.8 kc/s  & 8.3 kc/s    & 3.6\%      &   4.5\%   & 1.0 kbit/s        \\ [7pt] 
80 km  & 17.6 dB   & 0.031    & 0.015   &  1.7 kc/s  & 2.5 kc/s    & 5.1\%      &   5.9\%   & 0.14 kbit/s         \\ [7pt] 
    \hline\hline
\end{tabular}
\end{center}
\label{tab:exptab}
\end{table*}

\section*{Conclusion}
In this paper we propose the round-robin differential-phase-time-shift (RRDPTS) QKD protocol, in which two-bit key per state can be distributed, without modifying the experimental apparatus or relaxing the security assumptions of the original RRDPS protocol. The security proof of RRDPTS QKD under collective attacks has been derived. In addition, the performances of RRDPTS and RRDPS are compared for different experimental parameters. Our results show that the RRDPTS protocol with weak coherent pulses can work without monitoring signal disturbance. As compared with the RRDPS, the RRDPTS exhibits higher tolerance for interference errors, thus allowing for a higher secret key rate in the conditions of low visibility of interference. Furthermore, in order to test the performances our new QKD scheme in a more practical scenario, we reported a proof-of-concept experiment of one-decoy RRDPTS protocol over 80 km of single-mode fiber. 
This result paves the way towards a practical implementation of the round-robin QKD in real-world applications, in which the high error rate and channel instability are limiting the overall performance of current protocols.

\section*{Acknowledgments}
This work is supported by CSC Funding and by NSFC (Grant No.: 61831003), by the Center of Excellence SPOC - Silicon Photonics for Optical Communications (ref DNRF123), by the EraNET Cofund Initiatives QuantERA within the European Union’s Horizon 2020 research and innovation program grant agreement No. 731473 (project SQUARE), and by the NATO Science for Peace and Security program under Grant No. G5485.


\end{document}